\documentclass[nofootinbib,twocolumn,superscriptaddress,amsmath,amssymb]{revtex4}

\usepackage{amsmath}
\usepackage{bm}
\usepackage{siunitx}
\usepackage{graphicx}

\newcommand{\condcorr}[2]{\mathrm{Corr(#1\ |\ #2)}}
\newcommand{\normalcorr}[1]{\mathrm{Corr(#1)}}

\begin{document}

\title{Inferring simple but precise quantitative models of human oocyte and early embryo development}
\author{Brian D. Leahy}
\affiliation{Dept. of Molecular and Cellular Biology, Harvard University, Cambridge MA}
\affiliation{SEAS, Harvard University, Cambridge, MA}
\author{Catherine Racowsky}
\affiliation{Brigham Women's Hospital, Boston, MA}
\affiliation{Harvard Medical School, Boston, MA}
\author{Daniel Needleman}
\affiliation{Dept. of Molecular and Cellular Biology, Harvard University, Cambridge MA}
\affiliation{SEAS, Harvard University, Cambridge, MA}
\affiliation{Center for Computational Biology, Flatiron Institute, New York, NY}

\date{\today}

\begin{abstract}

Macroscopic, phenomenological models have proven useful as concise framings of our understandings in fields from statistical physics to economics to biology. Constructing a phenomenological model for development would provide a framework for understanding the complicated, regulatory nature of oogenesis and embryogenesis. Here, we use a data-driven approach to infer quantitative, precise models of human oocyte maturation and pre-implantation embryo development, by analyzing existing clinical In-Vitro Fertilization (IVF) data on 7,399 IVF cycles resulting in 57,827 embryos. Surprisingly, we find that both oocyte maturation and early embryo development are quantitatively described by simple models with minimal interactions. This simplicity suggests that oogenesis and embryogenesis are composed of modular processes that are relatively siloed from one another. In particular, our analysis provides strong evidence that (i) pre-antral follicles produce anti-M{\"u}llerian hormone independently of effects from other follicles, (ii) oocytes mature to metaphase-II independently of the woman's age, her BMI, and other factors, (iii) early embryo development is memoryless for the variables assessed here, in that the probability of an embryo transitioning from its current developmental stage to the next is independent of its previous stage. Our results both provide insight into the fundamentals of oogenesis and embryogenesis and have implications for the clinical practice of IVF.

\end{abstract}

\maketitle

\section{Introduction}
Understanding the manner by which a multicellular organism develops from a single cell is one of the grand challenges of biology. In mammals, this process begins with oogenesis inside the female, which results in an egg that becomes an embryo after fertilization. Early embryo development in mammals, including humans, is self-organized \cite{zhu2020principles, wennekamp2013self}: the course of events that unfold are governed by the embryo's internal dynamics and can proceed without external signals. Oogenesis and early embryogenesis have been studied from diverse perspectives, including molecular genetic, cell biological, chemical, and mechanical \cite{song2020chemical, gross2017active, chan2017coordination, fiorentino2020measuring, clift2013restarting, shahbazi2019self, rossant2009blastocyst, white2018instructions, hassold2001err, niakan2012human, jaffe2017regulation}. Despite the vast amount of knowledge that has been obtained, many basic questions remain, including: what determines which oocytes are selected for ovulation? How is the timing of embryonic events regulated? How are oogenesis and embryogenesis negatively impacted by age and disease? Answering these will provide fundamental insight and have strong implications for evolution and medical treatments of infertility. However, these issues are difficult to study using the molecular approaches that are the mainstay of current research, because of the integrated nature of the problems they pose concerning the overall trajectory of development.

An alternative to the microscopic, molecular perspective is to attempt to develop a macroscopic, phenomenological understanding. Such an approach has been productive in diverse areas from statistical physics~\cite{sethna2006statistical} to economics~\cite{bouchaud2003theory} to some fields of biology~\cite{bialek2012biophysics, needleman2017active}, including protein evolution~\cite{halabi2009protein} and cell-size control in bacteria~\cite{taheri2015cell, sauls2016adder, kohram2020bacterial}. One significant concern is that the great complexity of oogenesis and embryogenesis might make simple, phenomenological descriptions inapplicable. Furthermore, the validity of the phenomenological approach can only be determined by developing models and rigorously testing them. This requires a large amount of quantitative data, which is difficult to obtain from oocytes and embryos in model organisms.

Here, we overcome this challenge by leveraging a large data set from 7,399 routine clinical In-Vitro Fertilization (IVF) treatment cycles, resulting in 98,264 oocytes and 57,827 embryos. We show that the statistical structure present in this data can be quantitatively described using simple, phenomenological models. The models we develop are Bayesian networks, a form of probabilistic graphical model which represent conditional dependencies by a directed, acyclic graph. We infer models directly from the data, making little use of prior knowledge. Despite this, the resulting models recapitulate well-established aspects of oocyte and embryo development. Moreover, the resultant models are sparse: only one or two factors directly impact physiological processes. This implies that human oogenesis and embryogenesis are highly modular. Our analysis leads to a number of additional, surprising conclusions. We present strong evidence that:

i) each pre-antral follicle produces anti-M{\"u}llerian hormone (``AMH'') independently of effects from other follicles. This argues that AMH is a faithful indicator of the number of pre-antral follicles, consistent with its physiological role in regulating follicle recruitment and further supporting its clinical use as a measure of ovarian reserves~\cite{weenen2004anti, la2009anti, durlinger2002regulation, la2010anti}.

ii) while the number of oocytes released from follicles depends on many factors, the probability that a released oocyte matures to metaphase-II is independent of the patient's age, BMI, and other, external factors. This argues that physiological processes that are correlated with these external factors, such as mitochondrial metabolism and aneuploidy~\cite{franasiak2014nature,talmor2015female,broughton2017obesity}, do not significantly impact oocyte maturation.

iii) after oocytes are fertilized, the probability of successfully transitioning from one embryonic developmental stage to the next depends on the embryo’s present state, but not on its state at earlier stages. Thus, embryo development is Markovian, at least for the variables examined here. This argues that clinical embryo selection procedures need only consider the state of the embryos immediately before transfer, as the state of the embryo at earlier times provides no additional information.

Taken together, our results show that the development of oocytes and embryos emerges as a simple process, despite the underlying molecular complexities of the biology and despite the plethora of disease etiologies and treatment protocols presenting in a clinic. More broadly, this work validates the use of phenomenological models of oogenesis and embryogenesis by demonstrating that simple models can be constructed without sacrificing quantitative accuracy. Although we infer the models using data drawn from controlled ovarian stimulation and not from natural menstrual cycles, the models provide insight into the principles that govern oogenesis and embryogenesis, and may be useful in guiding clinical IVF treatments.

\section{Results}
\subsection{Oocyte Development}
\begin{figure*}
\includegraphics[width=\textwidth]{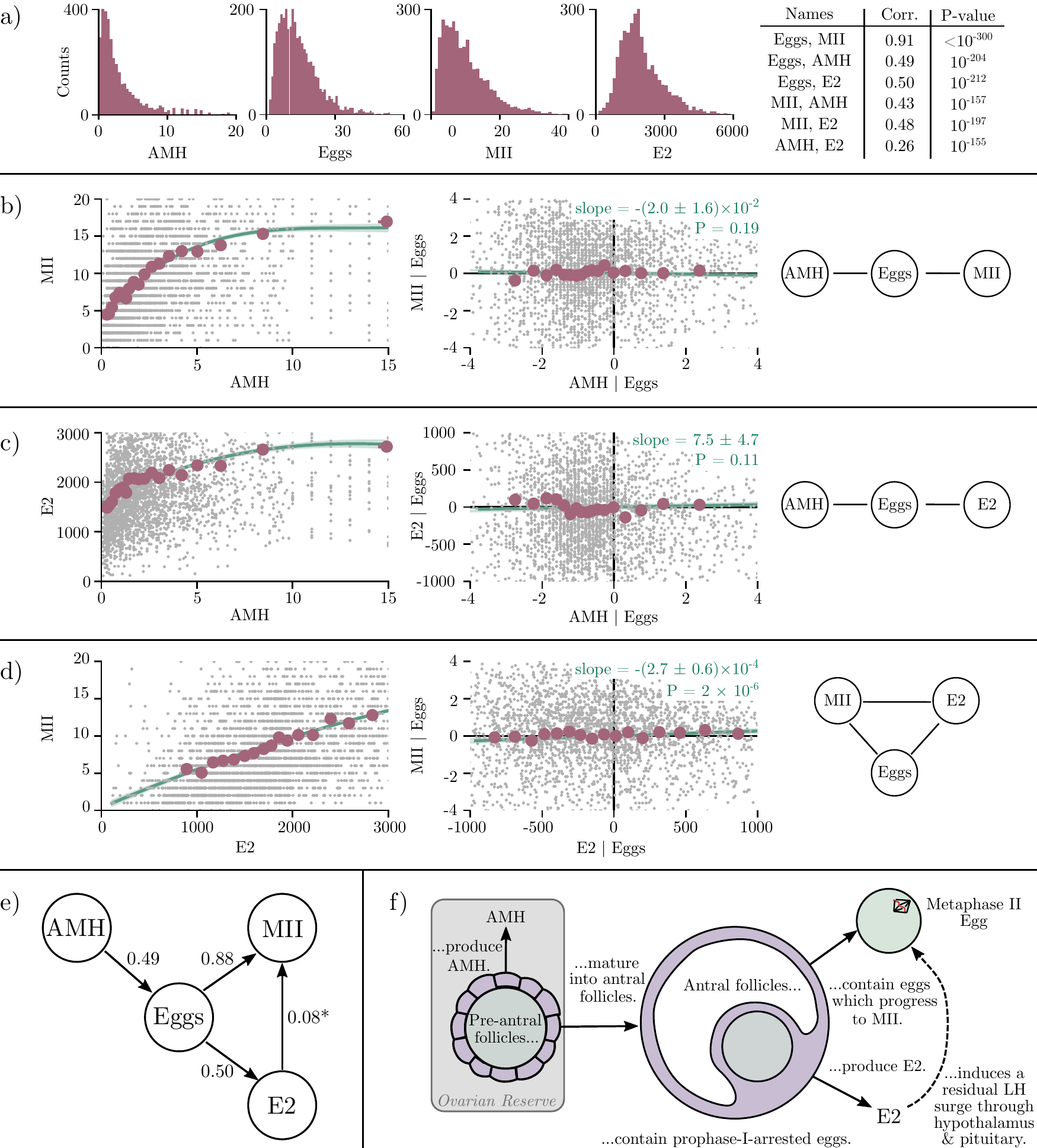}
\caption{
    (a) Distributions and correlations of the four variables AMH, Eggs,
    MII, and E2.
    (b) Left: The number of metaphase-II oocytes retrieved (MII) is
    strongly correlated with the patient's serum AMH. Gray dots: raw
    data, red circles: raw data binned into 20 separate bins with equal
    counts, green line and shaded region: nonlinear regression and
    errors. Center: That correlation disappears after regressing against
    the total number of retrieved oocytes (Eggs).  Gray dots: residuals
    after regressing against Eggs, red circles: residuals binned into 20
    separate bins with equal counts, green line and shaded region:
    linear fit to the residuals, with slope and standard error shown at
    top of plot. Right: This conditional independency suggests a graph
    of the form AMH -- Eggs -- MII.
    (c) Left: The patient's serum estradiol concentration (E2) is
    strongly correlated with AMH. Center: That correlation disappears
    after regressing against Eggs. Right: This conditional independency
    suggests a graph of the form AMH -- Eggs -- MII.
    (d) Left: MII is strongly correlated with E2. Center: While
    regressing against Eggs greatly weakens that correlation, E2 and MII
    remain correlated after conditioning on Eggs. This suggests a fully
    connected graph is needed to describe these three variables (right).
    (e) A graphical model that is consistent with the data. Edge labels
    show the conditional correlation coefficients after conditioning on
    all other incoming edges; the data is consistent with the arrow marked
    with a * oriented in either direction.
    (f) The graphical model expected from prior knowledge of ovarian
    stimulation.
    }
\label{fig:ovulation4element}
\end{figure*}

The formation of a healthy embryo depends on the successful progression of an oocyte from prophase-I arrest to metaphase-II. Thus, we begin by examining what determines the number of metaphase-II oocytes retrieved during an IVF ovarian stimulation. The clinical data contain 69 variables that describe either the patient or the ovarian stimulation. We start by examining four of the variables that are strongly correlated with the number of metaphase-II oocytes: (1) the number of total eggs retrieved during an ovarian stimulation (``Eggs''), (2) the number of eggs in metaphase-II arrest (``MII''), (3) the patient's maximum serum estradiol concentration during the treatment cycle (``E2''), and (4) the patient's serum anti-M{\"u}llerian hormone before the cycle (``AMH''). Estradiol is a hormone produced by the ovaries during both natural and stimulated ovulatory cycles~\cite{elder2020vitro}; AMH is considered a measure of the patient's ovarian reserve~\cite{la2010anti}. Each of these variables varies widely across the 4,910 cycles for which all four variables are recorded, with coefficients of variation of 0.5 -- 1.2 (Figure~\ref{fig:ovulation4element}a, left). All four variables are strongly correlated with one another, with correlation coefficients between 0.26 -- 0.91 and P-values between $10^{-300}$ -- $10^{-155}$ (Figure~\ref{fig:ovulation4element}a, right).

We start by searching for conditional independencies among the variables AMH, MII, and Eggs. A conditional independency between two variables implies that one variable can be completely described without direct knowledge of the other, suggesting that a simple, phenomenological model of the data exists. To search for conditional independencies, we nonlinearly regress MII on Eggs and AMH on Eggs, by finding the best-fit polynomial that maximizes the Bayesian posterior evidence. This method allows for capturing complex dependencies without overfitting the data~\cite{mackay2003information, mackay1992bayesian} (see SI Section 1); we also split the data into separate train and test sets as a further check against overfitting. We then take the residuals from the two regressions and evaluate their correlation. We denote this procedure as Corr(AMH, MII $|$ Eggs). We find that, although there is a strong correlation between AMH and MII (Figure~\ref{fig:ovulation4element}b, left), that correlation disappears after conditioning on Eggs: Corr(AMH, MII, $|$ Eggs) = -0.02 ($P=0.19$; Figure~\ref{fig:ovulation4element}b, center). This correlation is both consistent with zero and smaller than an effect size threshold of 0.05, suggesting that AMH and MII are conditionally independent given Eggs.

We encode this conditional independency using a class of graphical models known as Bayesian networks. These have found usage in causal inference~\cite{friedman2000using, pearl2009causality, buhlmann2014high, friedman2004inferring}; here, we use them to construct phenomenological models that correspond to mechanistic descriptions of biology. Briefly, for a given factorization of a probability distribution, these graphs contain a directed edge from one variable to another if the probability of the second variable depends on the first. Two variables are conditionally independent if all paths from one variable to the other are ``blocked'', where paths are blocked by head-to-tail or tail-to-tail nodes meeting at a variable that is conditioned on, or head-to-head nodes meeting at an edge that is not conditioned on~\cite{bishop2006pattern, pearl2009causality}. Both the observed correlation between MII and AMH and their conditional independency given Eggs can be captured by any of the graphs AMH $\rightarrow$ Eggs $\rightarrow$ MII, AMH $\leftarrow$ Eggs $\rightarrow$ MII, or AMH $\leftarrow$ Eggs $\leftarrow$ MII; we denote this ambiguity by AMH --- Eggs --- MII, with an as-yet undetermined orientation of the arrows (Figure~\ref{fig:ovulation4element}b, right). If the data are described by one of these graphs, then Eggs and AMH should remain correlated given MII, which is indeed the case: Corr(Eggs, AMH $|$ MII) = 0.24 ($P = 10^{-47}$; SI Figure~3a). Likewise, MII and Eggs should remain correlated given AMH, which is also the case: Corr(MII, Eggs $|$ AMH) = 0.87 ($P < 10^{-300}$; SI Figure~3b).

Next, we examine the variables AMH, E2, and Eggs. There is a strong association between AMH and E2 (Figure~\ref{fig:ovulation4element}c, left), but regressing both AMH and E2 on Eggs shows that Corr(AMH, E2 $|$ Eggs) = 0.03, which is consistent with no conditional correlation ($P = 0.11$; Figure~\ref{fig:ovulation4element}c, center). This suggests a graph of the form AMH --- Eggs --- E2 (Figure~\ref{fig:ovulation4element}c, right). Finally, we examine Eggs, MII, and E2. While regressing on Eggs greatly weakens the correlation between MII and E2 (compare Figure~\ref{fig:ovulation4element}d left and center), the measured conditional correlation is inconsistent with no correlation: Corr(E2,MII | Eggs) = 0.08, $P \approx 10^{-6}$. Thus, an edge must connect each of Eggs, E2, and MII (\ref{fig:ovulation4element}d, right).

Of the 543 graphical models that describe 4 variables, only 8 graphical models capture exactly the two conditional independencies in the data that are described above. The data alone cannot distinguish between these graphs. However, the patient's AMH is measured before the ovarian stimulation starts, whereas the other three variables are measured during the treatment. Thus, any graph with an edge pointing into AMH cannot correspond to a mechanistic description of the biology. Ruling these graphs out leaves only two graphs consistent with both the data and a mechanistic interpretation (Figure~\ref{fig:ovulation4element}e).

The quantitative phenomenological model encoded by these graphs recapitulates our qualitative understanding of ovarian stimulation (Figure~\ref{fig:ovulation4element}f): 1) Pre-antral follicles, which contain immature oocytes and associated somatic cells, produce the hormone AMH~\cite{weenen2004anti, durlinger2002regulation, pellatt2010anti}. Since pre-antral follicles have the potential to grow into large antral follicles with prophase-I-arrested eggs, AMH is a measure of the potential number of oocytes that could develop. This is captured by the inferred arrow in Figure~\ref{fig:ovulation4element}e from AMH to Eggs, which indicates that the patient's AMH determines how many eggs she will produce. 2) Antral follicles produce estradiol, captured by the inferred arrow from Eggs to E2. 3) Some, but not all, eggs progress from prophase-I arrest to metaphase-II arrest. This is captured by the inferred arrow from Eggs to MII. 4) During natural ovulation, the estradiol produced by antral follicles signals the pituitary and hypothalamus to release hormones which modulate oocyte maturation. In ovarian stimulation protocols, clinicians attempt to temporarily disable this feedback between the hypothalamus and the pituitary~\cite{elder2020vitro}, suggesting that estradiol should not impact oocyte maturation during ovarian stimulation. However, the inferred arrow from E2 to MII suggests that a weak feedback between the hypothalamus, the pituitary, and estradiol is still present during an ovarian stimulation cycle.

The inferred phenomenological model (Figure~\ref{fig:ovulation4element}e) provides a quantitative representation of oocyte development that allows direct and indirect effects to be disentangled. For instance, a patient starting treatment with a higher AMH is likely to produce more oocytes, via the direct arrow AMH $\rightarrow$ Eggs. In addition, that patient is likely to have a higher estradiol level during the cycle, as the additional eggs she is likely to produce will on average produce more estradiol, via the path AMH $\rightarrow$ Eggs $\rightarrow$ E2. However, the graph states that this effect is indirect: the patient’s E2 increases only through the associated increase in the number of eggs for high-AMH patients. This is borne out by the data. Likewise, a patient with a larger number of MII oocytes retrieved is likely to have a higher AMH, since following the arrows backwards shows that higher MII implies that Eggs is higher, and higher Eggs implies a higher AMH. However, once again, this is an indirect effect; the patient’s AMH is more likely to be high only because of the associated increase in Eggs when many MII oocytes are retrieved.

\subsection{Additional effects on Oocyte Development}
\begin{figure*}
\includegraphics[width=\textwidth]{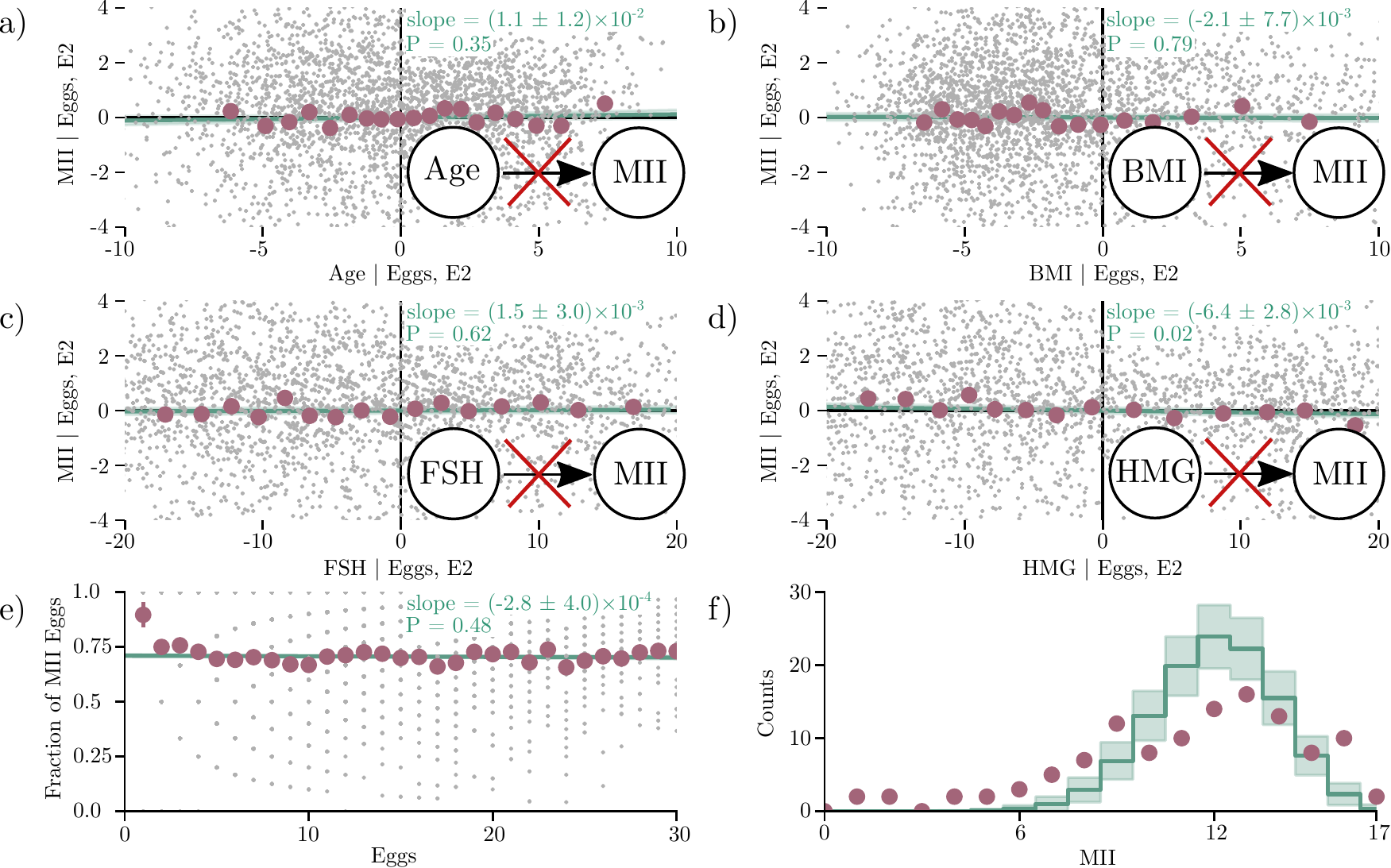}
\caption{
    MII versus the patient's age (a), BMI (b), and the doses of
    stimulation drugs FSH (c) and HMG (d), after regressing against Eggs
    and E2. Gray dots show the residuals from the regressions, red
    circles and error bars show the mean and standard error of the data
    binned into 20 bins with equal number of points, and green lines,
    shaded regions, and labeled slopes show the mean and standard error
    of the best linear fit to the residuals.
    (e) The fraction of MII oocytes (MII / Eggs) vs Eggs.
    (f) Histogram of observed MII (red circles) vs that expected
    from independently-triggering follicles (green line and shaded
    region show the expected counts and their standard deviation), for
    the 116 cycles that have 17 eggs retrieved, which is where the
    discrepancy between the two histograms is the largest as measured
    by a  $\chi^2$ test ($P < 10^{-300}$, primarily due to the cycles
    with MII of 1--5). On the scale of the plot, the expected histograms
    from a binomial distribution where the probability for an oocyte
    being metaphase-II is constant is indistinguishable from one where
    the probability varies with E2.
}
\label{fig:m2independence}
\end{figure*}

How do factors other than these four variables affect oocyte development? First, we examine the effect of the woman's age on an oocyte's ability to reach metaphase-II. The data show no conditional correlation between MII and the woman's age: Corr(Age, MII $|$ Eggs, E2) = 0.02 ($P = 0.35$, Figure~\ref{fig:m2independence}a). In addition, the data constrain the magnitude of any effect to be tiny. Fitting a line to the residuals constrains the slope to be $(1.1 \pm 1.2) \times 10^{-2}$ MII oocytes / year (point estimate $\pm$ standard error). To place this in perspective, consider a typical treatment cycle for a 33 year old and a 40 year old woman (the 25th and 75th percentile in the data). If the treatment results in the same number of total oocytes and the same max estradiol for both women, then on average the number of retrieved MII oocytes should differ by no more than 0.2. Since the median MII oocytes retrieved per cycle is 8, the data constrain the direct effect of age on MII to be 2\% or less. Next, we examine the effect of the woman's BMI on the oocyte. The data show no conditional correlation between MII and BMI: Corr(BMI, MII $|$ Eggs, E2) = -0.005 ($P = 0.78$; Figure~\ref{fig:m2independence}b). Moreover, the data constrain the direct effect of BMI on MII to be 1\% or less. Finally, the data show no conditional correlation between MII and the dose of either of the ovarian stimulation drugs FSH or HMG: Corr(FSH, MII $|$ Eggs, E2) = 0.01 ($P = 0.49$; Figure~\ref{fig:m2independence}c), Corr(HMG, MII $|$ Eggs, E2) = -0.03 ($P = 0.08$; Figure~\ref{fig:m2independence}d). Likewise, the data constrain the direct effect of these stimulation drugs to be 1\% or less for FSH, and 5\% or less for HMG. These observations show that oocytes develop to MII independently of a patient's age, her BMI, or details of the ovarian stimulation procedure.

Does the probability of an oocyte being in MII arrest depend on the total number of eggs retrieved? While Eggs is strongly predictive of MII, it provides no predictive power for the fraction of eggs in MII arrest (MII / Eggs): $\normalcorr{MII / Eggs, Eggs} = -0.01$ ($P=0.48$). Linearly regressing MII / Eggs on Eggs gives a slope tightly constrained near zero (Figure~\ref{fig:m2independence}e).

Combined, these observations suggest the following simple picture for oocyte maturation: Each follicle independently triggers its oocyte to leave prophase-I, with a probability that depends only on E2. The oocyte then progresses to metaphase-II, with both processes independent of interactions with other follicles or the aggressiveness of the ovarian stimulation.

To check whether this simple picture completely describes oocyte maturation, we examine the distribution of MII. If each follicle independently triggers its egg to progress to metaphase-II with a probability $p$, then MII for each cycle should be binomially distributed, denoted as $\mathrm{B}(\mathrm{MII}; \mathrm{Eggs}, p)$. If that probability depends only on E2, then for fixed Eggs the measured distribution of $\mathrm{MII}$ across cycles should be the average of many binomial distributions, each with a probability that depends on E2: $\langle \mathrm{B}(\mathrm{MII}; \mathrm{Eggs}, p(\mathrm{E2})) \rangle_\mathrm{E2}$. Instead, the empirical distribution of MII is much broader than the expected one (Figure~\ref{fig:m2independence}f). This discrepancy suggests that additional factors affect oocyte maturation. These additional factors could be biochemical processes within the oocyte that are shared by multiple eggs from the same patient, interactions between follicles beyond serum estradiol, or simply other clinical factors that we have not accounted for.

The data provide some insight into human meiosis, especially given prior knowledge of human ovulation. As a woman ages, the oocytes she ovulates become much more likely to be aneuploid, rising from an aneuploidy rate of roughly 25\% at age 30 to 80\% at age 42~\cite{franasiak2014nature, erickson1978down, morris2002revised}. Chromosomal signatures show that aneuploidy in human oocytes can arise both during the oocyte's progression from prophase-I to metaphase-II and immediately after fertilization~\cite{gruhn2019chromosome, hassold2001err, holubcova2015error, webster2017mechanisms}. In mitotic cells, mis-segregation of chromosomes is reduced by the spindle-assembly checkpoint~\cite{musacchio2015molecular}, which can arrest mitosis until chromosomes are correctly lined up. If there were a strong spindle assembly checkpoint in meiosis I, then the typically aneuploid oocytes from older women would reach metaphase-II at a reduced incidence than those from younger women. Instead, oocytes from older and younger women reach metaphase-II arrest at the same incidence. This is consistent with experimental work that shows that human oocytes have a weak meiotic spindle assembly checkpoint~\cite{mihajlovic2018segregating, holt2009control, howe2013recent}.

\begin{figure*}
\includegraphics[width=\textwidth]{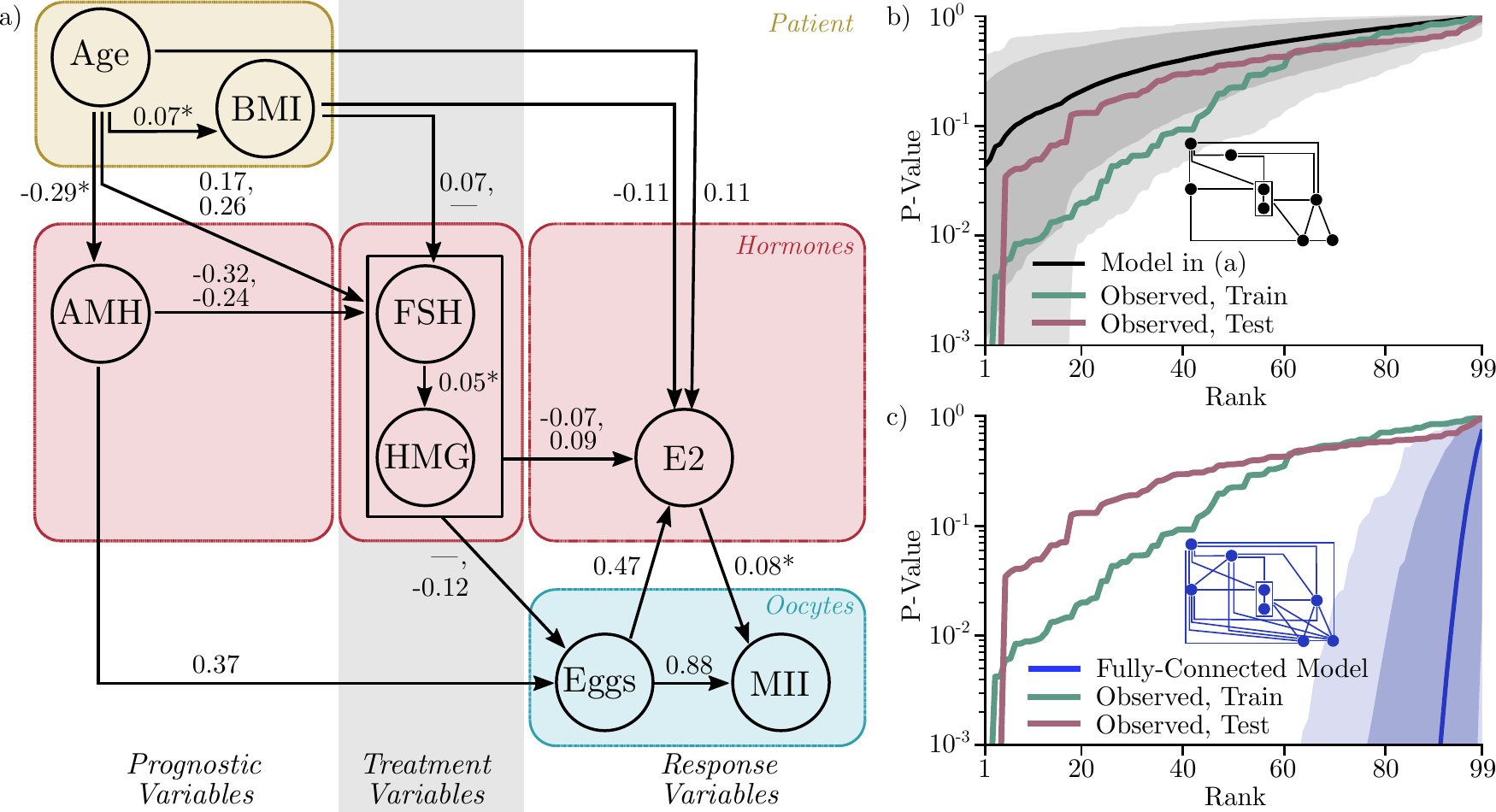}
\caption{
    (a) A graphical model of oogenesis that is consistent with
    a mechanistic interpretation of the data. Labels show
    conditional correlation coefficients. For edges with
    two labels, the upper corresponds to FSH, the lower to HMG;
    dashes signify no dependence. The data is consistent with
    marked arrows (*) oriented in either direction.
    (b) Rank plots of P-values for the 99 conditional correlations
    corresponding to the conditional independencies predicted by
    the model. The green line shows that from the training data, the red
    line that from the test data. The black line and shaded regions show
    the median, 95\%, and 99.9\% centered percentile of rank plots from
    3,000 datasets simulated according to the proposed model.
    (c) The same as (b), but showing the distribution of rank plots for
    the 99 conditional correlations from fully-connected, linear
    Gaussian models.
}
\label{fig:ovulation8element}
\end{figure*}

Next, we ask how the rest of oocyte recruitment is affected by the patient's age, BMI, and the doses of ovarian stimulation drugs FSH and HMG. We examine the joint distribution of all eight variables and construct a directed acyclic graph that is consistent with the data. The number of possible graphs grows rapidly with the number of variables: there are 543 possible graphs for four variables, but 783,702,329,343 possible graphs with eight variables. To deal with this inordinately large number of graphs, we use prior knowledge and split the variables into three groups: prognostic variables measured before the treatment starts (Age, BMI, and AMH), treatment variables (FSH and HMG), and response variables measured after the drugs have been applied (Eggs, MII, E2). We then search for graphs that are consistent with a mechanistic interpretation, by excluding graphs with edges directed from treatment to prognostic variables, from response to prognostic variables, or from response to treatment variables.

We start by examining the prognostic variables. There is one conditional independency among the prognostic variables: $\condcorr{AMH, BMI}{Age} = -0.04$ ($P=0.01$; SI Figure~4a). Physiologically, this implies that the likelihood that a primordial follicle develops into a pre-antral follicles is independent of any hormonal effects of obesity.

Next, we examine the treatment variables. The FSH dosage depends on all three prognostic variables, and the HMG dosage depends on Age, AMH, and FSH, but not BMI (SI Figure 4f). The complex dependencies of FSH and HMG on other variables imply that clinicians customize the treatment based on the patient, which is to be expected. Nevertheless, the lack of conditional independencies between the treatment variables and the prognostic variables demonstrates that the conditional independencies we do see elsewhere are real and not an artifact of our analysis technique.

Finally, we examine the response variables. After conditioning on AMH and HMG, Eggs is conditionally independent of both Age and BMI ($\condcorr{Age, Eggs}{AMH, HMG} = -0.04$, $P=0.01$; $\condcorr{BMI, Eggs}{AMH, HMG} = -0.01$, $P=0.54$; SI Figure~4c,d). Physiologically, this suggests that the ability of a pre-antral follicle to be recruited does not worsen with age or obesity. Likewise, Eggs is conditionally independent of FSH: $\condcorr{FSH, Eggs}{AMH, HMG} = -0.04$ ($P=0.02$; SI Figure~4e). This suggests that clinicians prescribe sufficient FSH to recruit all the follicles in the cohort activated from the primordial pool in that menstrual cycle. Interestingly, we observe a weak negative correlation between Eggs and the dose of HMG: $\condcorr{HMG, Eggs}{AMH} = -0.11$ ($P=10^{-11}$). Taken at face value, this implies that HMG is typically supplied at more than the optimal dose. While there is some evidence that excessive HMG can cause follicles to degrade~\cite{hugues2005does, wikland2001prospective}, another possibility is that the negative correlation is due to clinicians prescribing more HMG to patients whom they know \textit{a priori} to be poor responders even after accounting for their age, BMI, and AMH -- for instance, patients who have had a poor response in previous treatments. However, the correlation remains negative even when excluding patients on repeat stimulation cycles: $\condcorr{HMG, Eggs}{AMH,first~cycle}=-0.09$, $P \approx 10^{-5}$. Combined, the data paint a simple picture for oocyte recruitment: all available follicles are typically recruited, independently of effects from age or obesity but weakly affected by HMG. Finally, in contrast to the simplicity of Eggs and MII, E2 depends on all of Age, BMI, FSH, HMG, and Eggs. These combined observed conditional independencies are captured by the graphical model in Figure~\ref{fig:ovulation8element}a.

The model in Figure~\ref{fig:ovulation8element}a predicts 99 conditional independencies among the 8 variables shown. We measure the conditional correlations and associated P-values for each of these 99 independencies in both the train and test sets, and compare the distribution of these P-values to those calculated from 3,000 datasets simulated according to the model in Figure~\ref{fig:ovulation8element}a. The distribution of the P-values measured from the data is consistent with what is expected if the model is true, although the lowest P-value for the data is lower than that typical for the simulated data (Figure~\ref{fig:ovulation8element}b and SI Section~4). In contrast, datasets generated according to fully-connected models display a different distribution of these P-values (Figure~\ref{fig:ovulation8element}c and SI Section~4). Moreover, anything missing from the model in Figure~\ref{fig:ovulation8element}a must correspond to a small effect with small explanatory power. Fitting the training data with a fully-connected model explains 0.6\% or less of each variable's variance in the test data, with the fully-connected model actually performing worse on the test set for most of the fits than the model in Figure~\ref{fig:ovulation8element}a does. Combined, these observations show that the graphical model accurately describes human ovarian physiology and oogenesis.

Viewed holistically, the probabilistic graphical model in Figure~\ref{fig:ovulation8element} has a simple mechanistic interpretation. The patient's age and BMI only act to determine the hormone levels. Hormones other than estradiol determine how many antral follicles develop. The follicles then produce estrogen and trigger eggs to progress to metaphase-II, with a slight feedback between those two. Because of this simplicity, changes due to age or obesity manifest themselves in simple ways, after ignoring the physiologically irrelevant question of how clinicians choose drug doses for ovarian stimulation. In particular, the only direct effect of obesity on the oocyte maturation process is to decrease E2. This is consistent with work showing that obesity affects fertility by interfering with the hormonal regulation of ovulation~\cite{broughton2017obesity}. The primary effect of age on the oocyte maturation process is to decrease the ovarian reserve, as measured by the patient's AMH. Physiologically, this is consistent with the well-known decrease in a woman's pool of primordial follicles as she ages~\cite{elder2020vitro}.

\begin{figure}
\includegraphics{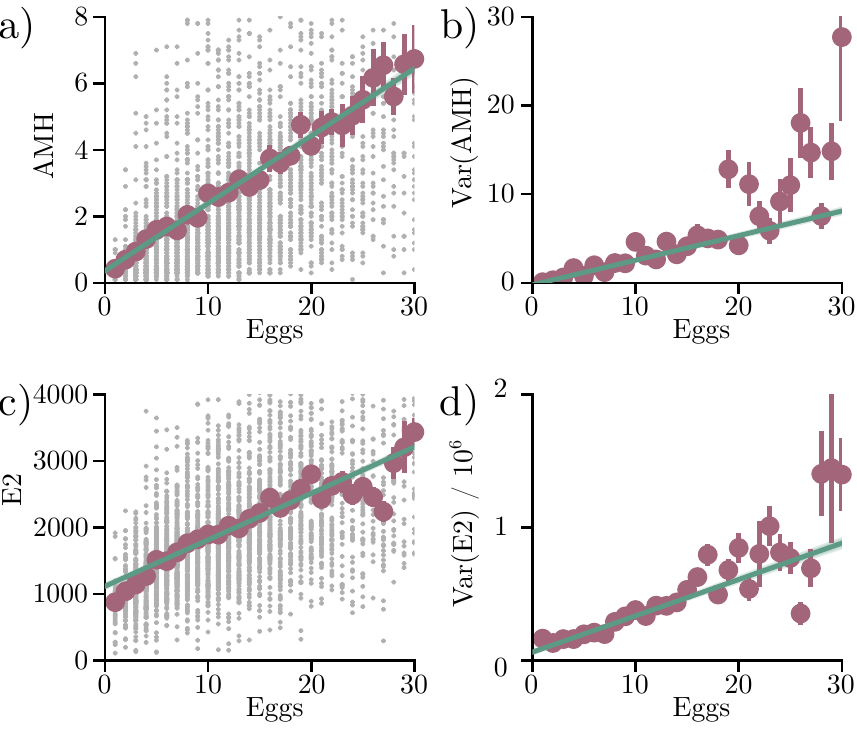}
\caption{
    (a) AMH vs Eggs. Gray dots show raw data, red circles show the mean
    and standard error of AMH binned at each value of Eggs, and green
    line shows the best linear fit to the data, with slope of
    $0.20 \pm 0.01$ and intercept $0.36 \pm 0.11$. A linear fit is the
    best-fit polynomial to the data, as determined by the model evidence.
    (b) Variance and error estimate of AMH vs Eggs, trimmed to
    the central 95\% for each value of Eggs (red dots, errors calculated
    using the variance of the k-statistic). The green line shows the
    best linear fit to the variance, with a slope of $0.28 \pm 0.01$ and
    intercept $-0.22 \pm 0.03$.
    (c) E2 vs Eggs. The green line shows the best linear fit to the
    data. A quadratic model (not shown) provides the best fit to E2 vs
    Eggs, as determined by the model evidence.
    (d) Variance and error estimate of E2 vs Eggs, trimmed to the
    central 95\% for each value of Eggs.
}
\label{fig:amhe2neggs}
\end{figure}

AMH and E2 are produced by ovarian follicles at different stages in follicular growth. The data provide insight into both these processes. For a given number of retrieved oocytes, the mean of AMH is linear in the number of retrieved oocytes, with an intercept near zero (Figure~\ref{fig:amhe2neggs}a, also SI Figure~5). For Eggs not too large, the variance of AMH is also linear in Eggs (Figure~\ref{fig:amhe2neggs}b); the deviation from linearity at large Eggs is due to patients with polycystic ovary syndrome, who tend to have high AMH. Since means and variances add when summing independent random variables, the linearity of both the mean and the variance of AMH in the number of follicles suggests that each follicle produces AMH independently from interactions with other follicles. (The small but nonzero intercept could arise if some pre-antral follicles do not mature sufficiently to be retrieved during the IVF retrieval procedure.) In contrast, the mean E2 is not linear in Eggs, systematically deviating from linearity and having an intercept that is far from zero (Figure~\ref{fig:amhe2neggs}c; variance in panel~d). These variations from linearity show that the E2 is not produced independently by each follicle, perhaps due to additional production from outside follicles (such as by adipose tissue or the adrenal glands), due to inter-follicle feedback, or simply due to external factors that affect E2 production, such as those shown in in Figure~\ref{fig:ovulation8element}.

\subsection{Pre-implantation development}

Once the oocyte is fertilized, the resulting embryo starts to divide. By the third day after fertilization, a human embryo typically has 8 cells. As the cells continue to divide, on the fifth day the embryo differentiates into a blastocyst, composed of two distinct cell lineages: the trophectoderm and the inner-cell mass. In natural fertilizations, the blastocyst then attaches to the woman's endometrial epithelium and implants in her uterus~\cite{cha2012mechanisms, elder2020vitro, niakan2012human, white2018instructions}. In the IVF clinic, human embryos are typically cultured for 3 or 5 days after fertilization, at which point embryologists attempt to select the highest quality embryo(s), which is then transferred into the patient's uterus.

To construct a phenomenological model of development, we first consider three variables that describe the overall trajectory of development: the embryo's number of cells on day 3 after fertilization (Day 3 Cells), its developmental stage on day 5 (Day 5 Stage), and whether it resulted in a fetal heartbeat after transfer (FH; see SI for details).

However, not all embryos are cultured to day 5, and thus not all embryos have data on both day 3 and day 5. Since the decision to culture embryos to day 5 is made based on patient prognosis and embryo quality, embryos that are assessed on day 5 systematically differ from those that are not. To avoid biases due to this missing data, we treat missingness as an additional variable and model both the variable and its missingness~\cite{little2019statistical}. The clinical data are in the “missing at random” regime, where a variable's missingness depends on other variables in the dataset, but not on the missing variable itself. In the language of probabilistic graphical models, there are edges from some of the normal variables to the missingness variables, but no edge from a variable to its own missingness. In this regime, valid inferences require conditioning on missingness and the variables on which the missingness depends. Multiple transfers cause an additional problem for the measurement of fetal heartbeat -- if two embryos are transferred simultaneously and one fetal heartbeat is observed, it is not obvious which embryo formed the fetus. We solve this problem via generative modeling. Briefly, we construct a parameterized model that predicts the probability of one embryo implanting from properties of the embryo and the woman. We then fit the model to the data by finding the maximum \textit{a posteriori} parameters, using a Poisson-binomial likelihood for multiple transfers. To check whether a variable is conditionally independent of fetal heartbeat, we fit two models, one with the additional variable and one without, and perform Bayesian model selection to see if the additional variable is necessary (SI Section~1). We include all available cycles with four or fewer embryos transferred (95\% of cycles).

\begin{figure*}
\includegraphics[width=\textwidth]{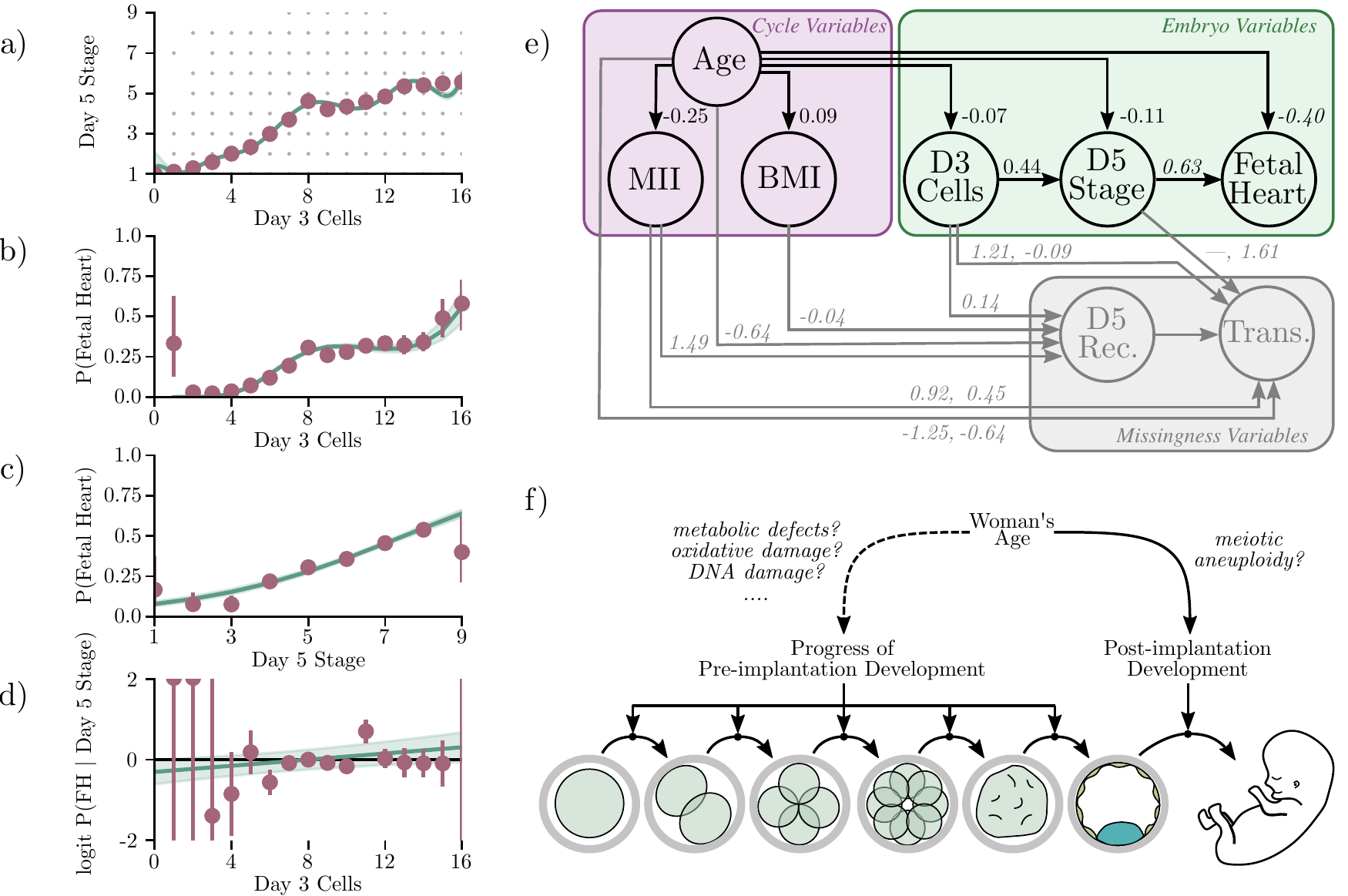}
\caption{
    (a) Day 5 Stage vs Day 3 Cells. Gray dots show the raw data, red
    circles show the mean Day 5 Stage for each separate value of Day 3
    Cells, the green line and shaded region show the nonlinear model
    with the highest evidence and its uncertainty.
    (b) Estimated probability of an embryo resulting in a fetal
    heartbeat (FH) as a function of Day 3 Cells alone, for embryos
    recorded on day 3 and transferred. The red circles and error bars
    show the probability estimated by a model that fits an independent
    probability of implantation for each number of cells; the green line
    and shaded region shows the nonlinear model with the highest model
    evidence and its uncertainty.
    (c) The estimated probability of FH as a function of Day 5 Stage
    alone, for embryos recorded on day 5 and transferred.
    (d) The logit of the estimated probability of FH as a function of
    Day 3 Cells, after regressing against Day 5 Stage. Red circles and
    error bars show the additional log probability estimated from a
    model that fits an independent logit for each value of Day 3 Cells;
    green line shows the best linear model and uncertainty for the
    logit. The data is consistent with Day 3 Cells having no additional
    predictive power on FH once Day 5 Stage is known.
    (e) The graph with the minimal number of edges that is consistent
    with the data and a mechanistic interpretation. Black nodes and
    arrows show the measured data; gray nodes and arrows show the data's
    missingness. Only the arrow Age $\rightarrow$ BMI can be re-oriented
    without breaking consistency with the data or a mechanistic
    interpretation. Edge labels for continuous variables are conditional
    correlation coefficients (roman typeface).  Edge labels for discrete
    variables are coefficients from logistic regression (italic
    typeface), after treating the effects of other variables with edges
    into the discrete variable and after normalizing the input variable
    by its mean and standard deviation.  The missingness variables D5
    Rec. and Trans. are 1 if the embryo is recorded on day 5 or
    transferred, respectively, and 0 if that information is missing. The
    distribution of Trans. changes depending on whether Day 5 Stage was
    recorded; the two labels on edges into Trans correspond to Day 5
    Stage missing or recorded.
    (f) The data is consistent with a picture where processes which
    control pre-implantation development are largely different from
    those which control post-implantation development.
}
\label{fig:development}
\end{figure*}

We use this approach to construct a model of human pre-implantation development. The embryo's number of cells on day 3 is strongly correlated with its stage on day 5: $\condcorr{Day\ 3\ Cells, Day\ 5\ Stage}{both\ measured, Age, BMI, MII} = 0.44$ ($P<10^{-300}$, Figure~\ref{fig:development}a). The embryo's Day 3 Cells is also predictive of fetal heartbeat (Figure~\ref{fig:development}b), as is appreciated in the literature~\cite{racowsky2011national, vernon2011utility}. Likewise, the embryo's Day 5 Stage is predictive of fetal heartbeat (Figure~\ref{fig:development}c)~\cite{gardner2016assessment}. However, the probability that a transferred embryo results in a fetal heartbeat depends only on its stage on day 5; Day 3 Cells provides no additional information whether the embryo will develop (Figure~\ref{fig:development}d). This conditional independence implies a model of the form Day 3 Cells $\rightarrow$ Day 5 Stage $\rightarrow$ FH; this is the only graph with two or fewer edges that is consistent with the data and the fact that day 3 happens before day 5.

To understand how properties of the woman and her ovarian stimulation affect embryo development, we examine the effect of three other variables: the woman's age, her BMI, and the number of MII oocytes retrieved in the ovarian stimulation. The woman's age is weakly correlated with the embryo's number of cells on day 3 and its stage on day 5: $\normalcorr{Day~3~Cells, Age} = -0.07$ ($P=10^{-47}$; SI Figure~6a), $\condcorr{Day~5~Stage, Age}{Day~3~Cells, measured} = -0.11$ ($P=10^{-78}$; SI Figure~6b). In contrast, the woman's age has a much stronger effect on the probability that an embryo forms a fetal heartbeat (SI Figure~7c). Surprisingly, the patient's BMI is conditionally uncorrelated with all of Day 3 Cells (SI Figure~6c), Day 5 Stage (SI Figure~6d), and FH (SI Figure~7e). Likewise, the conditional correlation of MII with all of D3 Cells, D5 Stage, and FH is either consistent with zero or less than 0.05 in magnitude (SI Figure~6e,f and 7f). Combined, these observations yield the simple graphical model for development in Figure~\ref{fig:development}e. The woman's age determines the embryo's cell number on day 3; her age and the embryo's cell number determine the embryo's stage on day 5; and her age and the embryo's stage determine whether it will continue to develop. Neither the patient's BMI nor the number of retrieved MII eggs significantly affect the embryo's development. In contrast, the data's missingness shows a much more complex distribution, with almost all variables affecting the clinical decisions regarding embryo transfer and culture duration. Once again, the data paint a picture of simple biology but complex clinical decisions.

The lack of a directed edge from the number of cells on day 3 to fetal heartbeat is particularly striking. The time-varying morphology (morphokinetics) of human embryos is known to be predictive of developmental success~\cite{elder2020vitro, vernon2011utility}. In principle, morphokinetics up to day 3 could provide a different set of information than morphokinetics between days 3 and 5. For example, since the embryo's genome activates on day 3~\cite{braude1988human}, one might reasonably propose that the embryo's progress before day 3 provides information about the oocyte's cytoplasm, that the embryo's progress from day 3 to day 5 provides information about aneuploidy, and that both of these factors independently determine the embryo's prognosis. Instead, the data show that the embryo's stage at day 5 contains all the information about its viability that its stage at day 3 contains.

One physiological interpretation consistent with the data is that there are two distinct sets of mechanisms that influence an embryo's development (Figure~\ref{fig:development}f). One set of mechanisms influences pre-implantation development only through its overall rate, including the number of cells on day 3 and the stage on day 5. This set of mechanisms depends weakly on age. Another set of mechanisms influences an embryo's post-implantation developmental potential and depends strongly on age. Perhaps surprisingly, meiotic aneuploidy of the oocyte cannot strongly affect pre-implantation development, since meiotic aneuploidy is strongly associated with the woman's age. Instead, the data suggest that meiotic aneuploidy primarily affects post-implantation development, rather than pre-implantation development. Other works have provided mixed evidence whether aneuploidy affects pre-implantation development~\cite{rienzi2015no, minasi2016correlation, campbell2013modelling, kaser2014clinical}.

To see how robust this picture is, we augment the analysis with additional information on the embryo's pre-implantation development. In addition to the number of cells, the dataset also describes the embryo on Day 3 with the presence of cytoplasmic fragments, the presence of multiple nuclei in individual cells, size asymmetries among cells within the embryo, the presence of large vacuoles, and the granularity of the cell cytoplasm. On day 5, the dataset also includes grades of the inner-cell mass and the trophectoderm of embryos that have formed blastocysts. Of the day 3 variables, embryo fragmentation and cell symmetry are predictive of fetal heartbeat, along with Age and Day 3 Cells. Likewise, of the day 5 variables, the trophectoderm grade is predictive of fetal heartbeat, along with Age and Day 5 Stage, in agreement with recent studies~\cite{hill2013trophectoderm, ahlstrom2011trophectoderm}. (The data are consistent with the other day 3 variables providing no additional predictive power once Age, D3 Cells, symmetry, and fragmentation are known; and the data are consistent with the inner-cell mass grade providing no additional predictive power once Age, D5 stage, and the trophectoderm grade are known.) Nevertheless, the day 3 variables provide no additional predictive power for fetal heartbeat once the day 5 variables are known (SI Section~4). These observations suggest a memoryless model of pre-implantation development: provided the embryo makes it to the blastocyst stage, what happened before is irrelevant for its viability. Thus, despite the molecular complexities of early development and the complicated trajectory of human development before 12 weeks, a simple, coarse-grained view of embryonic viability may be possible without sacrificing quantitative accuracy.

\section{Discussion}

Here, we have used clinical IVF data and minimal prior knowledge to infer quantitative, phenomenological models of human oogenesis and embryogenesis. Not only does constructing these models with a data-driven approach give confidence in their validity, but the models recapitulate known aspects of oogenesis and embryogenesis as well. Surprisingly, the models that best describe the data are sparse, with only one or two factors affecting most physiological processes. This suggests that oogenesis and embryogenesis are modular processes. Moreover, our analysis leads to three additional, surprising conclusions which support this overall picture of modularity.

i) AMH production by one pre-antral follicle is independent of the amount produced by others. This is in stark contrast to other hormones produced by the follicle. Hormones such as estradiol and inhibin participate in feedback loops which regulate the formation of the dominant follicle in a natural cycle. As a result, these hormones are produced in a highly regulated manner, and not independently by each follicle~\cite{elder2020vitro,macklon2006science}. Moreover, mathematical modeling suggests that, for these feedback loops to function, the hormone production and response needs to be highly nonlinear~\cite{akin1984ovulation, lacker1981regulation, lacker1988ovaries}. Thus, the amount of these hormones produced by one follicle depends strongly on the amount produced by other follicles. In contrast, AMH appears to be produced without regulatory feedback. This is particularly interesting because AMH also regulates the number of growing follicles in a cohort, by regulating the growth of primordial follicles into primary (pre-antral) follicles. Thus, while feedback loops are needed to accurately control the number of mature, Graafian follicles recruited during natural ovulation, feedback loops appear to be unnecessary to sufficiently control the number of primary follicles recruited.

ii) Neither age, obesity, the ovarian stimulation, nor even the number of recruited oocytes affects whether an individual oocyte progresses to metaphase-II once triggered to resume meiosis. These observations have several implications for the biology of the oocyte. Definitive evidence shows that age is strongly correlated with aneuploidy in the oocyte~\cite{franasiak2014nature,erickson1978down}. However, since our analysis shows that the ability of an oocyte to progress to metaphase-II is independent of age, we conclude that this ability is the same for both euploid and aneuploid oocytes. Thus, the spindle assembly checkpoint in human oocytes must be weak, in agreement with recent experimental work~\cite{mihajlovic2018segregating, holt2009control, howe2013recent}. A similar argument can be made regarding the effect of metabolism on meiosis. Evidence suggests that oocyte mitochondrial metabolism worsens with increasing obesity~\cite{leary2015human, broughton2017obesity}. Since oocytes progress to metaphase-II independently of obesity, metabolic defects must not typically be enough to stop an oocyte from progressing from prophase-I to metaphase-II.

iii) Early embryonic development is memoryless, in that embryos with the same status on day 5 develop the same, regardless of their status on day 3. This memorylessness is reminiscent of the robustness of early mammalian embryos to damage to individual cells~\cite{van2008four, willadsen1980viability, allen1984production}; however, memorylessness is more than robustness. Robustness signifies that an embryo can recover from a setback. Memorylessness signifies that, once recovered, neither the setback nor what caused it has any impact on the rest of development. The memorylessness implies a modularity in development~\cite{hartwell1999molecular}.

The models we present also have implications for clinical IVF. For embryos transferred on day-5, embryo selection can be based solely on how developed they are on day 5, independent of their status on day 3. While day 3 information is correlated with implantation potential, the effect is completely captured by the embryo's status on day 5. For ovarian stimulation, the data provide no evidence that aggressive ovarian stimulation is detrimental to the oocyte, either in its ability to mature to metaphase II, to develop as an embryo, or, if transferred, to form a viable pregnancy (Figures~\ref{fig:ovulation8element} \& \ref{fig:development}); moreover, the data constrain any of these effects to be small. Thus, a clinic should not be concerned about a potential tradeoff between the quality and quantity of retrieved oocytes. Conversely, the data show that, at the clinic from which our data were derived, the ovarian stimulation drugs FSH and HMG are typically applied at saturating or slightly deleterious doses for follicular recruitment. Thus, the hormone dosage could presumably be slightly reduced, to mitigate side effects such as ovarian hyper-stimulation syndrome or the high cost of stimulation drugs, without a large decrease in the number of retrieved oocytes.

More broadly, our results provide fundamental insights into the overall process of development. Unlike the sparseness that arises in theoretically-motivated models simply as a way to manage complexity, the sparseness in our models is a property of the data. This sparseness implies that the biology itself is simple, consisting of modularized processes that are quantitatively siloed from one another. The simplicity is particularly surprising given the many ways that oogenesis and embryogenesis are affected by diseases, including age-related infertility, endometriosis, sperm malfunction, and polycystic ovary syndrome, all of which are present in our dataset. Overall, our results suggest that, despite their underlying complexities, oogenesis and embryogenesis are modular processes that result in simple, emergent behavior, and that a concise, quantitative understanding of the rest of development may be possible.

\section{Acknowledgments}
We would like to acknowledge K. Walden, S. Kou, V. Manoharan, and the Needleman lab for useful discussions. BL acknowledges the Harvard Quantitative Biology Initiative for support and useful discussions. DN and CR acknowledge the National Institutes of Health for financial support (R01HD092550-01).

\bibliography{bibliography.bib}

\end{document}